# Energy minimization and AC demagnetization in a nanomagnet array

X. Ke, J. Li, C. Nisoli, Paul E. Lammert, W. McConville, R. F. Wang, V. H. Crespi, and P. Schiffer

Department of Physics and Materials Research Institute, Pennsylvania State University, University Park, PA 16802, USA

We study AC demagnetization in frustrated arrays of single-domain ferromagnetic islands, exhaustively resolving every (Ising-like) magnetic degree of freedom in the systems. Although the net moment of the arrays is brought near zero by a protocol with sufficiently small step size, the final magnetostatic energy of the demagnetized array continues to decrease for finer-stepped protocols and does not extrapolate to the ground state energy. The resulting complex disordered magnetic state can be described by a maximum-entropy ensemble constrained to satisfy just nearest-neighbor correlations.

The generation of a well-defined thermodynamic "state" is fundamental to the statistical analysis of physical systems. Classical magnetostatic systems, with their well-defined interactions between elements, provide an important test-bed for such analyses, whether as continua (Landau-Ginzburg), discretized sets of continuous degrees of freedom (Heisenberg, XY), or discretized sets of two-level systems (Ising). Inducing a statistically well-defined state in these systems can be problematic, since the domain structure of ferromagnetic samples and the moment arrangement of magnetic island arrays are often controlled by their history of exposure to magnetic fields. For bulk ferromagnetic samples, domain structure is typically modified via AC demagnetization, whereby an alternating external field of gradually shrinking strength carves a disordered arrangement of mutually compensating domains of variable size and orientation. AC demagnetization has also been applied to obtain an initial demagnetized state in regular arrays of nanomagnets and thin films [1,2,3,4,5,6]. Theoretically, such AC demagnetization allows energy minimization in special cases such as Preisach models of ferromagnets [7] and certain disordered systems [8,9,10]. On the other hand, there has been little quantitative experimental study of how demagnetization relates to magnetostatic energy minimization, or its impact on local domain structure or island moment correlations.

A nanoscale single-domain magnet island array, wherein the Ising-like moment of every island is experimentally resolvable, allows detailed study of the relation between AC demagnetization and magnetostatic energy across different demagnetization protocols. The ability to place Ising-like degrees of freedom with fixed, controllable, relative orientation and spacing affords a great degree of control in designing new physics into such arrays. Magnetic island arrays can be engineered to minimize island-island interactions [11] for data storage applications, or to balance shape anisotropy against array anisotropy [12]. Here we use such arrays to study the process of AC demagnetization and demonstrate that minimizing the overall system moment does not define a unique minimum energy state. For successively more finely stepped demagnetization protocols, the total magnetostatic energy of our arrays continues to decrease long after the overall magnetization has been minimized, and ground-state-like correlations continuously strengthen with decreasing step size, without limiting to the ground state energy. Monte Carlo modeling indicates that the resulting collective magnetic state can, in fact, be characterized by a well-defined thermodynamic ensemble in which entropy is maximized subject to an explicit constraint on the nearest-neighbor correlations.

We study lithographically fabricated "artificial spin ice" arrays of single-domain ferromagnetic permalloy islands (Fig. 1), with the island moments constrained to point along their long axes by a strong shape anisotropy [5]. The arrays have a range of lattice constants for the underlying square lattice (400 nm to 880 nm), a fixed island size (220 nm x 80 nm and 25 nm thick), and a coercive field $H_c$ ~ 770 Oe [4] that is nearly independent of lattice spacing. Since the islands are single-domain, magnetostatic coupling between individual Ising-like island moments controls the system's energetics in zero applied field [1,2,3,13,14]. Dipolar interactions between the four islands at each cross-like vertex are frustrated, i.e. they cannot all be simultaneously satisfied. However, the system has a simple two-fold degenerate ground state with zero net moment, as shown in Fig 1(a). Since the island-island interaction energy and the intra-island shape anisotropy energy are large compared to accessible thermal energies, AC demagnetization is required to obtain a demagnetized state. Previous work by our group showed that such demagnetization can produce a disordered low-moment state without long-range correlations [4,5].

We subjected the arrays to AC demagnetization by rotating them about an axis perpendicular to the sample plane at 1000 rpm in a stepwise decreasing in-plane magnetic field. The field magnitude was reduced from 1280 Oe to 0 with constant size steps, $\Delta H$, and the field direction reversed at each step, as shown in Fig. 1(c). We chose $\Delta H$ = 1.6, 3.2, 9.6, 12.8, 16, 32, or 64 Oe, holding each step for 5 seconds and ramping the field at 24,000 Oe/s between steps. After each demagnetization protocol, magnetic force microscopy (MFM) images of the moment configurations (Fig 1(b)) were acquired at several locations on each array, well away from the array boundaries and totaling about 5000 islands for each case. The uncertainty of quantities derived from the images is calculated as the standard deviation of the individual image results.

As the initial field of 1280 Oe is much larger than both $H_c$ and the inter-island fields (which are ~10 Oe [5]), the samples begin in a fully magnetized state with both horizontal and vertical island moments fully aligned. The island moments presumably track the applied field as long as it significantly exceeds $H_c$, the island-island interactions being small perturbations. The magnetostatic interactions between islands most strongly influence their spin orientations in a narrow window around $H_c$, and the island moments become fixed once the external field drops well below $H_c$. For the largest step size, $\Delta H$ = 64 Oe, the arrays are polarized to a certain direction when the field is larger than $H_c$, and then become largely fixed in that configuration as the field takes a single step to a value well below $H_c$, yielding a net magnetized state. In contrast, at smaller $\Delta H$, the island moments freeze into a more random configuration associated with variations in local magnetic field due to inter-island interactions.

We now quantitatively analyze how the collective state of the moments depends on $\Delta H$. To assess the effectiveness of the demagnetization, we measure the residual net magnetic moment of the arrays, $m_{tot} = \sqrt{m_x^2 + m_y^2}/\sqrt{2}$ [4] where $m_x$, $m_y$ are the net normalized moments along Cartesian directions. As shown in Fig. 2, the residual moment is large for the larger $\Delta H$, but it is statistically indistinguishable from zero [15] when $\Delta H$ is smaller than 12.8 Oe, a threshold corresponding approximately to the strength of the magnetic fields exerted by one island on another. To test whether the window near the coercive field is truly crucial to demagnetization, we altered the protocol so that $\Delta H$ differed within a field window around $H_c$ (from 992 Oe to 554 Oe). For $\Delta H = 32$ Oe outside the window and $\Delta H = 3.2$ Oe within, the final results are indistinguishable from a protocol with a uniform $\Delta H = 3.2$ Oe. Similarly, for $\Delta H = 3.2$ Oe outside the window and $\Delta H = 32$ Oe within, the results are indistinguishable from a protocol with only $\Delta H = 32$ Oe.

For any step size smaller than 12.8 Oe, the final state is effectively demagnetized. But is the demagnetized state for $\Delta H < 12.8$ Oe truly independent of $\Delta H$? To answer this question, we examine correlations between nearby moment pairs of three types, as shown in Fig. 1(d): Longitudinal (separated parallel to their long axes), Transverse (separated perpendicular to their long axes) and Diagonal (separated along a diagonal of the lattice). We define the correlation between these moments as +1 (or -1) when an island pair follows (or opposes) the ground-state pair correlation. The average correlations are labeled $L(1)...L(5)$, $T(1)...T(5)$, $D(1)...D(5)$ out to 5$^{th}$ nearest neighbors. The two left-hand columns of Fig. 3 show the experimental correlations for the 400 nm lattice at step sizes of 1.6 Oe and 12.8 Oe. The small increase in $D(n)$ and $T(n)$ as $\Delta H$ drops from 12.8 Oe to 1.6 Oe reflects gradually increasing correlation lengths. The $L(n)$ correlations show a more dramatic change, rising from nearly zero for $\Delta H$=12.8 Oe to a clear development of correlations at $\Delta H$ = 1.6 Oe. Since $D(1)$ and $T(1)$ show well-developed correlation even for the larger $\Delta H$, the lack of $L(n)$ correlation is not simply the result of random non-interacting islands. Rather, it reflects the frustrated nature of the system, since the direct pair-wise interaction favors $L(1) = -1$, whereas the indirect influence mediated through other islands favors the ground state with $L(1) = 1$. (Both D(1) and L(1) pairings prefer head-to-tail alignments, but these are not mutually compatible: in the ground state of Fig. 1(a), the stronger interaction, D(1), wins, so L(1)=1 defines a head-to-head or tail-to-tail pairing.) The right-hand column of Fig. 3 shows three near-neighbor correlations as a function of $\Delta H$. These correlations continue to evolve down to the smallest $\Delta H$=1.6 Oe step, long after the net magnetization has essentially disappeared.

Clearly, knowledge of the net magnetization alone does not uniquely specify the statistical nature of the array's state, since many states of equivalent magnetization have very different local correlations. We now demonstrate that this complex non-equilibrium demagnetized state (or at least all pair-wise correlations within it) can be described by just two numbers: $D(1)$ and $L(1)$. This surprising result is demonstrated by the Monte Carlo calculation of Fig. 3 (red dashed curve). The simulation samples the most random ensemble that satisfies a specified $D(1)$ and $L(1)$- see Supplemental Information for computational details. The correlations of the Monte Carlo simulation for a 400 nm lattice are in excellent agreement with the experimental data; the comparison is equally good for other lattice parameters. Therefore, all distance-dependent *L, D* and *T* correlations are simply consequences of a given *L(1)* and *D(1)*, when imposed on a state of maximum information entropy [16]. Since such a state for an entire lattice is formally identical to a canonical state with nearest-neighbor interactions that generates the desired correlations, the Monte Carlo approach is justified.

Even though these magnetic correlations continue to evolve within the demagnetized state, the system does *not* approach the ground state in the limit $\Delta H \rightarrow 0$. A careful convergence study indicates that summing all pair-wise island interactions out to $7^{th}$ neighbors (using micromagnetics calculations [17] to determine the magnetization within a single island) provides a good approximation to the total magnetic interaction energy of the arrays, since they are overall demagnetized and two-dimensional systems have weak contributions from far neighbors. The upper panel of Fig. 4 plots the total magnetostatic interaction energy of the 560 nm array as a function of $\Delta H$ summed to $7^{th}$ neighbor, plus the individual contributions from just the *D(1)*, *L(1)* and *T(1)* pairs (and the sum of these three). As anticipated, the frustrated *L(1)* contribution to the energy actually increases with decreasing step size, especially below $\Delta H = 12.8$ Oe where the overall magnetization is minimized; only the compensating decreases in *D(1)* and *T(1)* enable the total energy to fall, and the three-neighbor summation of *D(1)*, *L(1)*, and *T(1)* (green dashed line) is already indistinguishable from the total energy. The lower panel of Fig. 4 shows the evolution of the magnetic interaction energy as a function of field step size for all the lattice constants studied, normalized so that the ideal ground state energy is -1. The total magnetostatic energy of the array continues to decrease with decreasing $\Delta H$, and the slope of $E(\Delta H)$ is not noticeably altered after the moment is minimized at $\Delta H \leq 12.8$ Oe. The ideal ground state energy is well below the $\Delta H \rightarrow 0$ limit of the data for all lattice constants. Furthermore, the normalized limiting energy decreases with decreasing lattice spacing: stronger interactions (at a given field step size) facilitate energy minimization.

Since the array energy extrapolates to values well above the ground state energy as the step size $\Delta H \rightarrow 0$, our results strongly suggest that AC demagnetization alone cannot access the ground state, even well past the point where the overall system moment is minimized. Physically, this is allowable because a broad manifold of microstates with zero net moment and strong local correlations minimize the interaction locally; such manifolds are characteristic of frustrated [18] or jammed [19] systems with multiple competing interactions. Nevertheless, a thermodynamic specification of this complex state is still possible, in terms of only two nearest-neighbor pair-wise correlations. While our study has focused on a specific island lattice, the process of AC demagnetization is relevant to a broad range of magnetic systems, and it has analogs in many athermal systems such as vibration-induced packing of granular materials [20]. The unusual degree of control provided by interacting nanomagnet arrays thus provides an important tool in the investigation of how athermal systems can be moved towards lower-energy states. For example, the transition between jammed and unjammed magnetic lattices could be mapped out as a function of island spacing and topology while simultaneously monitoring all relevant dynamical degrees of freedom, perhaps revealing e.g., analogs of fragile and strong glass formers in the magnetic domain.

We acknowledge support from Army Research Office, the National Science Foundation MRSEC program (DMR-0213623), and the National Nanotechnology Infrastructure Network. We are grateful to Prof. Chris Leighton for providing the samples and for helpful discussion.

**FIGURE CAPTIONS**

**Figure 1:** The experimental system under study. (a) Island array with the moment configuration in its ground state; (b) MFM image of arrays with 680 nm lattice spacing at $\Delta H$ = 12.8 Oe. The black and white dots correspond to the south and north poles of the island moments, respectively, and the AFM image inset shows the array lattice. The scale bar is 1 µm; (c) Schematic of the AC demagnetization protocol applied while the sample was rotated within the field (not to scale). The inset shows a schematic of the demagnetization setup; (d) Near-neighbor pairs $D(n)$, $L(n)$, and $T(n)$ as represented by dark blue, dark red, and light green islands, respectively, relative to the centered grey island.

**Figure 2:** Field step size dependence of residual moment as defined in the text. Inset is an expanded view for the region of low step size.

**Figure 3: Left Two Columns:** correlations between island moments, referenced to the ground state for Diagonal, Longitudinal and Transverse relative positions, as defined in Fig. 1, for the 400nm lattice at field steps of 1.6 and 12.8 Oe. Black curves give experimental measurements and red dashed curves give Monte Carlo results as described in the text. **Right Column:** Step-size dependence of pair-wise correlations $D(1)$, $L(1)$, and $T(1)$ for 400 and 880 nm lattice spacing. All symmetry-equivalent D, T and L directions are included in the computations.

**Figure 4: (a)** Step-size dependence of the total magnetostatic energy of just *D(1), L(1),* or *T(1)* island pairs, the sum of these three, and the complete sum out to seventh neighbors for the array at 560 nm lattice spacing. **(b)** Normalized total energy for an array as a function of field step size for arrays with 400 nm, 560 nm, 680 nm, and 880 nm lattice spacings. The black square corresponds to the energy of the ideal ground state of Fig. 1(a).

Figure 1
X. Ke, et al

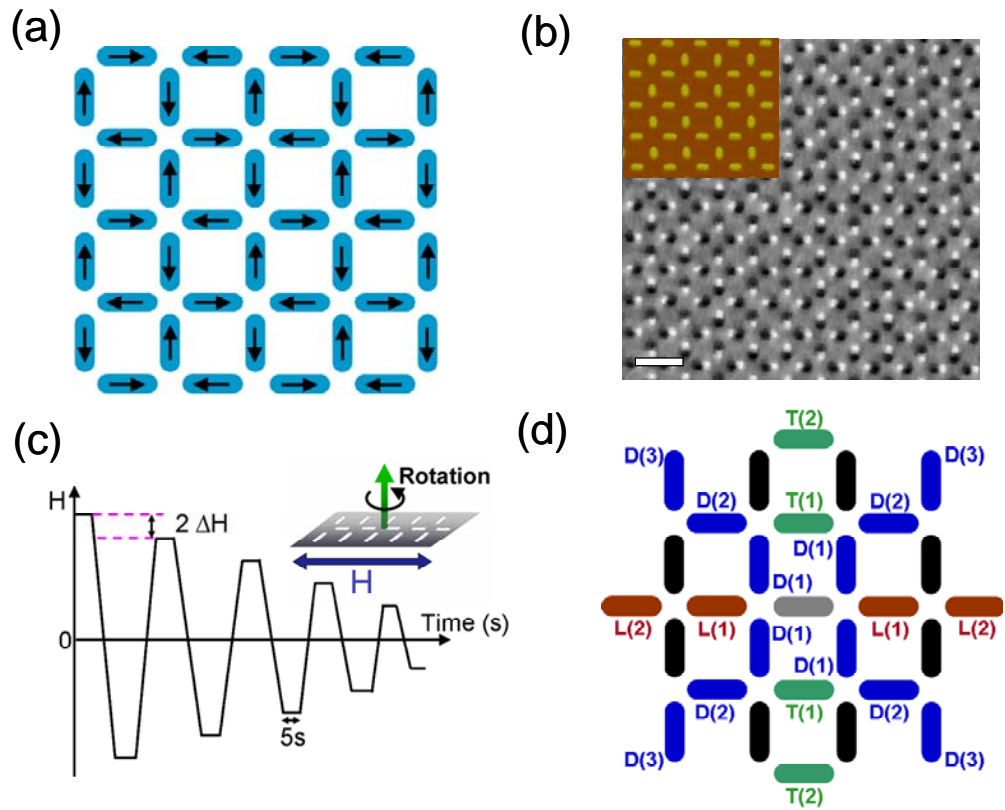

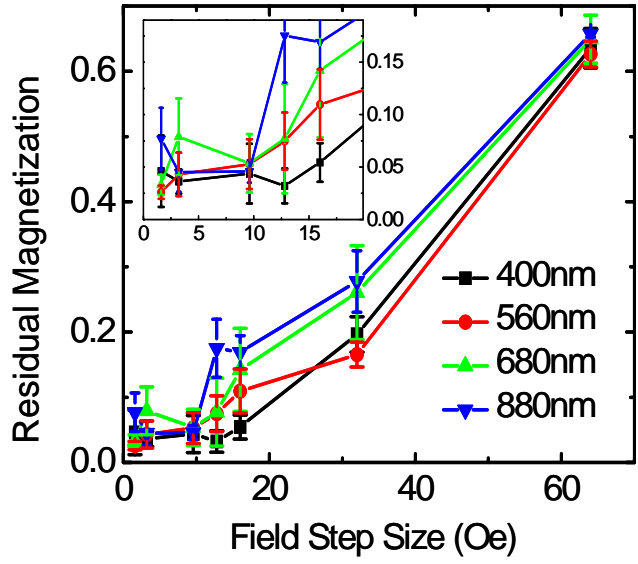

Figure 3.
X. Ke, et al

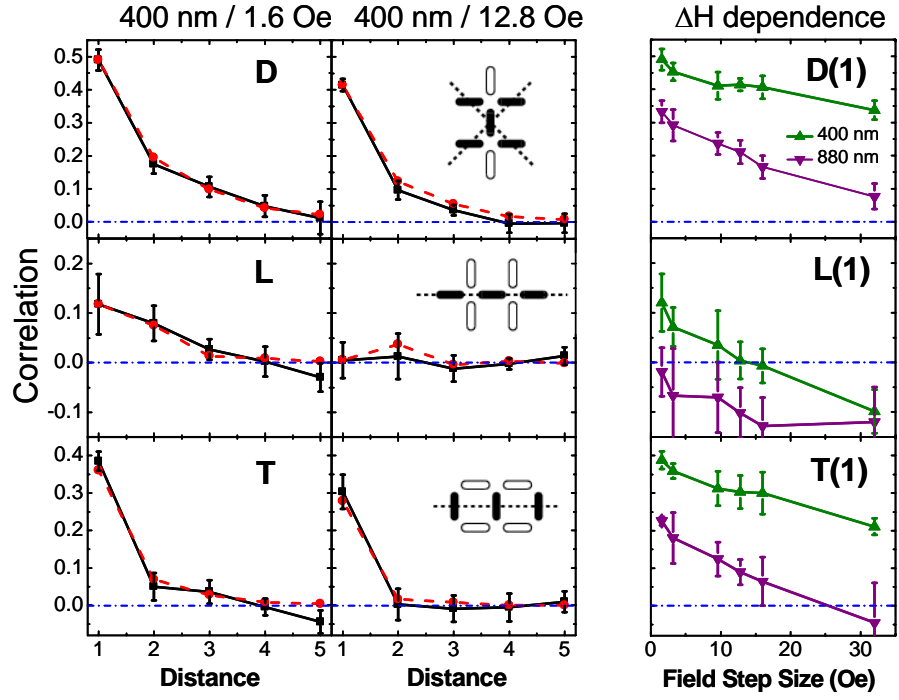

Figure 4.
X. Ke, et al

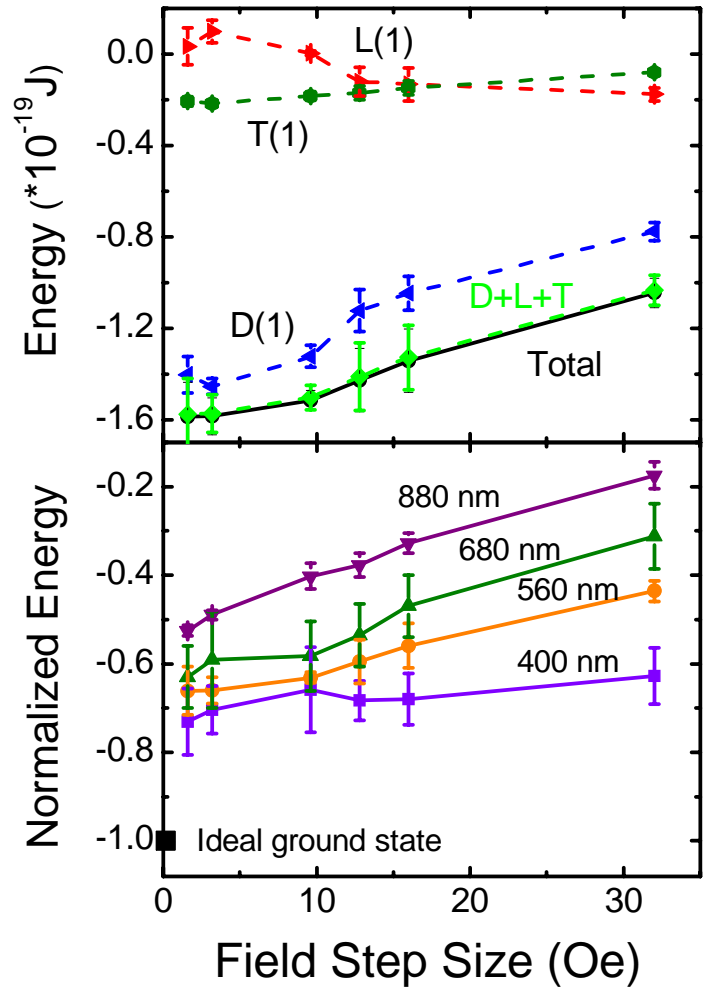